\documentclass[%
 reprint,
 amsmath,amssymb,
 aps,
floatfix,
]{revtex4-2}

\usepackage{graphicx}
\usepackage{color}
\usepackage{dcolumn}
\usepackage{bm}

\usepackage[ruled, linesnumbered]{algorithm2e}

\usepackage{xcolor}
\DeclareMathOperator{\lw}{\text{lw}}

\usepackage{mathtools}
\usepackage{physics} 
\DeclarePairedDelimiter\ceil{\lceil}{\rceil}

\usepackage{siunitx}

\begin{document}

\title{A local pre-decoder to reduce the bandwidth and latency of quantum error correction}

\author{Samuel C. Smith}
 \affiliation{Centre for Engineered Quantum Systems, School of Physics, University of Sydney, Sydney, NSW 2006, Australia}

\author{Benjamin J. Brown}%
 \affiliation{Centre for Engineered Quantum Systems, School of Physics, University of Sydney, Sydney, NSW 2006, Australia}

\author{Stephen D. Bartlett}%

 \affiliation{Centre for Engineered Quantum Systems, School of Physics, University of Sydney, Sydney, NSW 2006, Australia}

\date{\today}

\begin{abstract}
A fault-tolerant quantum computer will be supported by a classical decoding system interfacing with quantum hardware to perform quantum error correction. It is important that the decoder can keep pace with the quantum clock speed, within the limitations on communication that are imposed by the physical architecture. To this end we propose a local `pre-decoder', which makes greedy corrections to reduce the amount of syndrome data sent to a standard matching decoder. We study these classical overheads for the surface code under a phenomenological phase-flip noise model with imperfect measurements. We find substantial improvements in the runtime of the global decoder and the communication bandwidth by using the pre-decoder. For instance, to achieve a logical failure probability of \(f = 10^{-15}\) using qubits with physical error rate \(p = 10^{-3}\) and a distance $d=22$ code, we find that the bandwidth cost is reduced by a factor of 1000, and the time taken by a matching decoder is sped up by a factor of 200. To achieve this target failure probability, the pre-decoding approach requires a 50\% increase in the qubit count compared with the optimal decoder. 

\end{abstract}

\maketitle

\section{Introduction} 

Performing complex calculations on a quantum computer will require maintaining quantum information coherently for long periods of time.  To achieve this with noisy devices, quantum error correction (QEC) can be used, wherein measurements of code stabilizers throughout the computation give rise to a syndrome that can be interpreted by decoding software, allowing errors to be tracked and corrected~\cite{ReliableQuantumComputers, ReviewQECforQM}. Current demonstrations of real-time error correction~\cite{RealTimeProcessing,HoneywellRealTime, QECGoogleAI, IBMRealTime, ETHZurich} are constrained by several pragmatic concerns. First, the classical decoder must keep pace with the high rate at which syndrome data is produced. In addition, the amount of syndrome measurement data transferred between the quantum layer and the decoding unit may be limited by some maximum bandwidth set by device constraints. These constraints are sufficiently stringent that it can be difficult to find a software-based solution to accelerate classical decoding such that it runs at the speed that is demanded by the quantum hardware.

To accelerate the decoding problem and reduce the communication costs, we develop a two-level decoding scheme for the surface code~\cite{KitaevToricCode, TopologicalQuantumMemory} involving a local pre-decoder based on cellular automata (CA)~\cite{CAforTQM, FTCAforTQM, Gacs, HarringtonThesis, CAforNonAbelianAnyons, BiasedCA, TopologicalQuantumMemory, SweepRule, SweepRuleAndBeyond, LocalPerformanceAnalysis}. Our decoding scheme is designed to correct sparsely distributed errors by making greedy decoding decisions at the pre-decoder stage. The pre-decoder corrects these simple errors, however, its local nature means that locations where multiple errors have occurred may remain uncorrected. The pre-decoder sends the updated syndrome after it applies its correction to a minimum-weight perfect matching (MWPM) decoder to complete the decoding process. 

As the CA can be implemented in parallel on dedicated hardware, without requiring long-range communication with a central processing unit, we find that our implementation has a significant impact on the latency and bandwidth requirements of the global decoding problem. Specifically, by reducing the number of errors, and in turn the number of syndrome defects, global decoding~\cite{TopologicalQuantumMemory, MWPMPhenomenologicalThresholdBitFlip, RGforQudits, MWPMINOrder1, ExpandingDiamonds, CubicCodesHardRG, ImprovedHDRG, UnionFind, RateAgnosticMCMC, DecodingPerspective, NeuralBeliefPropagation, NeuralDecoder, Varsamopoulos2017Neural, LinearTimeGeneralDecoder, FastDecodersSoftRG, HighThresholdMCMCDecoder, RateAgnosticMCMC, Bravyi2014EfficientMaxLikelihood, ProjectionDecoder, RestrictionDecoder, MobiusStrip} is simplified and can therefore be completed at a higher speed. Moreover, by reducing the number of syndrome defects per unit volume, i.e., the defect density, we can reduce the size of the message that needs to be passed to the global decoder using a suitable strategy for data compression.

Previous and concurrent work has considered using the full hardware stack to distribute and pipeline the QEC workload~\cite{TamingInstructionBandwidth, ScalableDecoderMicroarchitecture, ChamberlandFastLocalDecoders, Ueno2022, Gicev2021, Meinerz2022, HierarchicalDecoding, Ravi2022, LILLIPUT, Paler2022, CAforTQM, FTCAforTQM,  HarringtonThesis, QECOOL, AQEC}. Although some processing of the syndrome measurement data can be carried out adjacent to the quantum layer, the amount of processing that can be done is subject to device constraints such as thermal budgets. Within this setting, several proposals have been made for two-level decoding schemes, including approaches where the first stage has been implemented either by sophisticated neural networks~\cite{ChamberlandFastLocalDecoders, Ueno2022, Gicev2021, Meinerz2022}, or by lightweight decoders that correct the easy error configurations~\cite{HierarchicalDecoding, Ravi2022, LILLIPUT, Paler2022}. There also exist decoding schemes that attempt to correct all errors using on-chip cryogenic hardware at some expense to the accuracy of the decoder \cite{QECOOL, AQEC, QULATIS}.

In our approach, our local pre-decoder will always attempt to correct the error, and is designed to be accurate wherever the error is sufficiently sparse. For almost all error rates below threshold, up to \(p < 1.8\%\), this results in a reduction to the defect density of the syndrome, with the defect density scaling quadratically in the error rate and a 1000X reduction being attained at error rates \(p=10^{-3}\). The corresponding runtime of the main MWPM decoder is made to scale quartically in the error rate, instead of quadratically as for the case without pre-decoding. In the limit of large surface code patches and at error rate \(p = 10^{-3}\), a resulting speed up of a factor of 1900 is possible, with a 200X speed up realised at this error rate for the \(d=22\) code. In assessing the cost of the scheme, entropic factors make a significant contribution to the magnitude of failure rates~\cite{RoleOfEntropyInTQEC}, particularly in practical regimes. We find that even though our decoding scheme performs sub optimally in the large system-size limit, by careful consideration of entropic factors we find that the tradeoff we propose are less costly for systems that we expect may be experimentally accessible in the near term. These explain the modest 50\% increase in the qubit count that is required over a wide range of error rates and code distances, as well as the smaller increase of 15\% required when \(d < 9\).

The structure of this paper is as follows. In Section~\ref{sec:decoding}, we discuss fault-tolerant decoding on the surface code and define the error model that we will use. In Section~\ref{sec:predecoding}, we develop the pre-decoding algorithm and its action on the syndrome history. In Section~\ref{sec:failure-rates-and-threshold}, we demonstrate evidence of a threshold and study sub-threshold failure probabilities. In Section~\ref{sec:savings} we present numerical results on the bandwidth reduction that is observed, and on a corresponding saving in the runtime of a MWPM main decoder. In Section \ref{sec:practical-cost}, we discuss the performance of the pre-decoder over a practical regime. We identify combinatoric and finite-size effects that are important to consider and provide evidence that the asymptotic scaling of failure probabilities is a pessimistic estimator of performance in this regime. Finally, in Section~\ref{sec:discussion}, we conclude and discuss future work. In Appendix~\ref{sec:recovering-failure-rates}, we generalize the pre-decoder to a parameterized family of pre-decoders, allowing a smaller logical failure probability to be recovered by trading off some of the bandwidth usage and latency savings.

\section{Fault-tolerant error correction on the surface code \label{sec:decoding}} 

\begin{figure} 
    \includegraphics[width=0.7\columnwidth]{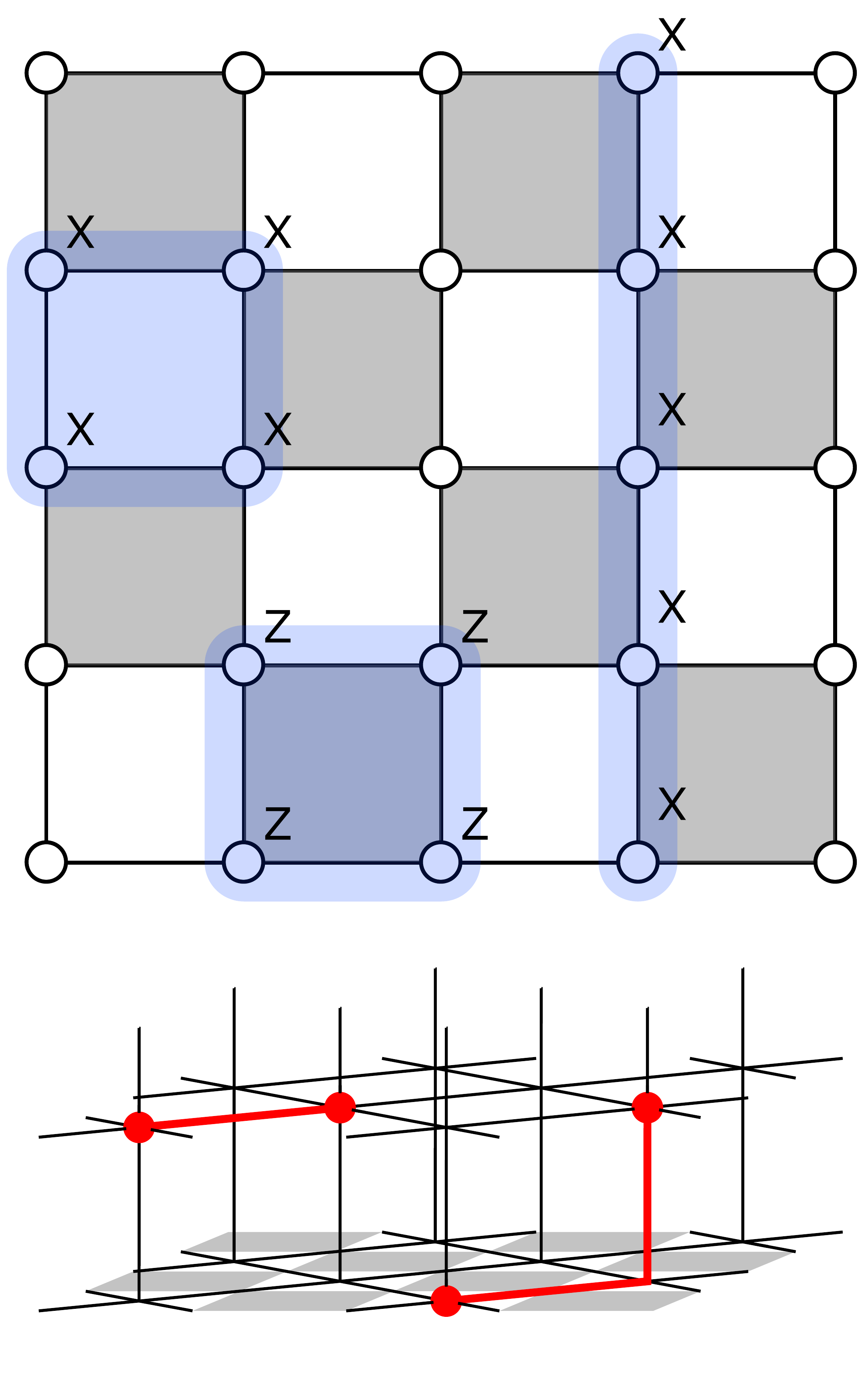}
    \caption{\label{fig:surface-code} (top) The distance-4 rotated surface code. Qubits are denoted by white circles. Periodic boundary conditions are enforced by identifying qubits along the top row and the bottom row, as well as along the left column and the right column. Pauli \(X\)-type (\(Z\)-type) stabilizers are given by the product of Pauli \(X\)-type (\(Z\)-type) operators over white (shaded) squares. One stabilizer of each Pauli type is shown. The surface code encodes two logical qubits into the shared \((+1)\)-eigenspace of the stabilizers. Logicals operators are given by the product of Pauli operators around a non-trivial cycle, and a Pauli \(X\)-type logical operator is shown. Logical operators commute with all the stabilizers, implying that they have a non-trivial action that preserves the logical code space. (bottom) The decoding lattice used for decoding phase-flip noise and faulty \(X\)-type stabilizer measurements. Only two rounds of stabilizer measurement are shown, although for a distance-4 code, four rounds of stabilizer measurement would be required. Space-like edges correspond to possible locations of phase-flip errors, and time-like edges correspond to possible locations of measurement errors. When an error occurs on an edge, whether time-like or space-like, a pair of defects is created on the vertices contained in the boundary of the edge. A single phase-flip error is shown, as well as an extended string-like error consisting of a phase-flip error and a measurement error. The string-like error creates defects at its endpoints.}
\end{figure}

Our scheme can be defined over any topological code with arbitrary boundaries. For concreteness, we choose here to focus on the rotated surface code with periodic boundaries (see Fig.~\ref{fig:surface-code}). Since the surface code is of Calderbank-Shor Steane (CSS) type, we can deal with the correction procedure for Pauli-Z and Pauli-X errors separately. We will concentrate on Pauli-Z type errors, but remark that an equivalent discussion will hold for the Pauli-X errors due to the duality of the different types of stabilizer. 

We demonstrate our  decoding algorithm using an independent and identically distributed (i.i.d.)\ noise model that introduces phase-flip errors, and where stabilizer measurements may also give unreliable outcomes. This simplified noise model captures many of the phenomenological features of the gate-based noise model that reflects the physics of locally interacting hardware. We assume that all of the stabilizer generators of the code are measured simultaneously at a fixed frequency, whose clock cycle defines a natural unit of time. All qubits experience an i.i.d.\ dephasing noise channel, where the channel will introduce a Pauli \(Z\) error to a qubit with probability \(p\) per unit time, otherwise the qubit experiences no error. Further, we assume that the stabilizer measurements are imperfect, and that each measurement returns an incorrect outcome with probability \(p_m\), and we set \(p_m = p\).

In order to protect encoded information we need to perform active error correction.  As we are considering faulty measurements, we cannot rely on any individual measurement outcome. The standard approach, which we employ here, is to repeat each stabilizer measurement \(d\) times in order to build up a (2+1)-dimensional set of outcomes referred to as the \emph{syndrome history}. The syndrome history will serve as the input to the decoding algorithm, and an example syndrome history is shown in Fig.~\ref{fig:surface-code}. A syndrome history variable is specified by an index \(v\) over \(X\)-type stabilizers, and a time \(t\), and is denoted \(s_v(t)\). In particular, \(s_v(t)\) is defined as the parity of measurement outcomes of the \(X\)-type stabilizer associated to the index \(v\) taken at times \(t, t-1\). Phase flip and measurement errors correspond to space-like and time-like edges in the syndrome history respectively, and string-like error configurations create defects at their endpoints. 

A decoder takes this syndrome history and proposes a correction that will recover the encoded state with high probability. That is, a decoder is an algorithm whose input is a (2+1)-dimensional syndrome history and whose output is a Pauli operator that will return the code to a code state. We remark that with this definition, a decoder is not concerned with the possibility that a string of measurement errors will wrap non-trivially through the syndrome history in the time-like direction. The decoder fails when the net result of the error and the correction does not return the system to the same state that it started in, and an important quantity of interest for a decoder is its failure probability with respect to a given noise model.

\section{Decoding and pre-decoding}
\label{sec:predecoding}

In this work we propose the use of local pre-decoder to reduce the bandwidth demands and latency for a global decoding algorithm, specifically, the MWPM decoder. We first briefly review the well-studied matching decoder, before introducing our pre-decoder.

\subsection{Minimum-weight perfect matching}

Decoding the surface code for a local noise model involves pairing nearby defects of the syndrome. This problem is particularly well suited for MWPM~\cite{TopologicalQuantumMemory}; see Refs.~\cite{TowardsPracticalClassicalProcessing, DecodingPerspective} for a review.

A perfect matching is a subgraph of some input graph such that each vertex of the output subgraph has exactly one incident edge. Efficient algorithms are known to find a MWPM for a graph, where the input graph has weighted edges, and the output graph is such that the sum of the weights of its edges is minimal~\cite{Edmonds1, Edmonds2}. 

It is well-known that an algorithm for MWPM can be used to correct the surface code reliably~\cite{TopologicalQuantumMemory}. In order to do so,
we create the complete graph whose vertices correspond to defects in the syndrome history. Weights are assigned to edges that connect defect pairs in proportion to the probability that the defect pairs were created at the endpoints of one string-like error. For an i.i.d. noise model together with measurement errors at the same rate \(p_m = p\), we have the simplification that weights are calculated from the taxicab metric on the syndrome history. The output of the MWPM algorithm is a choice of edges that matches each vertex to exactly one other vertex. From the output matching, a string-like Pauli operator can be computed that will restore the system to the code state. 

Assuming the MWPM algorithm has prior information about the error model, the decoder requires a list of defects from the $(2+1)$-dimensional syndrome history. The latency and the bandwidth requirements depend on the number of defects from this volume, as follows. Firstly, the runtime of the MWPM decoder scales polynomially with the number of vertices of its input graph, which we denote by \(\abs{W}\). For instance, here we use an implementation with a typical runtime scaling like \(O(\abs{W}^2)\). Second, the bandwidth requirements for communicating the syndrome to the decoder may also depend on the number of defects.  Although sending the full record of all syndrome measurements depends only on the spacetime volume $V$ of the syndrome history, this is very inefficient in the limit that the error rate is very small such that defects are sparsely distributed. Such a syndrome history can be compressed such that the length of the message is proportional to the number of defects produced by an error configuration, simply by sending a list of addresses where defects are identified. Given that we can describe the address of a defect in the spacetime volume $V$ with $\log V$ bits, we have that the syndrome can be communicated with a message of total length $O(\bar{\rho} V \log V)$ on average, where \(\bar{\rho}\) is the mean density of defects in the syndrome history and we anticipate that $ \bar{\rho} = 2 p$. (Note that fluctuations around this average message size need to be considered, and we point the reader to Ref.~\cite{ScalableDecoderMicroarchitecture} where this and other strategies for syndrome compression are considered.)

As we have argued, we can improve the latency and the bandwidth required to use a global decoder by minimising the number of defect locations that are communicated beyond the quantum device. The goal of a predecoding is to attempt to correct small errors locally, and only to pass difficult error cases to the global decoder.

\subsection{Pre-decoding}

The pre-decoder is the first stage of a pipelined, two-stage decoding scheme. It can be understood as a map on syndrome histories that returns a modified syndrome history and a partial correction. The modified syndrome history then gets passed to a global decoder which returns a correction. The total correction for the code is given by combining the partial correction returned by the pre-decoder and the correction returned by the global decoder. Although it is useful to think of the pre-decoder as a map on syndrome histories, it is also important to keep in mind that it is implemented locally in space and time by CA that can sit adjacent to the quantum layer. Additionally, the CA and global decoder can both operate concurrently, with the pre-decoder cleaning up the syndrome history as stabilizer measurements are taking place, and at the same time the global decoder is working on processing the syndrome history from the previous error correction round. 

The pre-decoder is based on the greedy matching of nearest-neighbour defect pairs to reduce the defect density. Such an approach works well when combined with syndrome compression techniques to ensure that the bandwidth scales with the number of defects in the code. We remark that due to the ease and parallelisable nature of nearest-neighbour matchings, they have attracted some attention recently. For example, the pre-decoder can be viewed as a truncated version of a HDRG decoder~\cite{CubicCodesHardRG}, and a similar algorithm has been used to compute correlations amongst errors for MWPM~\cite{Paler2022}. 

We now define the pre-decoder precisely. At each edge on the decoding lattice (space-like and time-like) we define an update rule that determines whether a matching should be made on that edge. In particular, a matching will be made if and only if the boundary of the edge is contained inside the syndrome. If the edge is time-like, the defects on the boundary are removed from the syndrome and no further action is taken. If the edge is space-like, the defects are removed from the syndrome and a phase-flip is recorded. It is only the parity of phase-flips that needs to be updated here, so only a single bit of local memory is required. 

As described thus far, these update rules are not well-defined because we have not specified the order in which these matchings get carried out. To resolve this, we have that the matchings get carried out in one concurrent step. A matching at any edge in the syndrome history is always determined by the syndrome data prior to pre-decoding in the boundary of that edge. More formally, the rules define a map on syndrome histories that outputs a modified syndrome and a partial correction. The modified syndrome is given: 

\begin{equation} 
    s_v'(t) = s_v(t) + s_v(t) \Bigl( \sum_{u \sim v} s_u(t) + s_v(t + 1) + s_v(t - 1) \Bigr). 
    \label{eq:modified-syn} 
\end{equation} 

The partial correction returned by the pre-decoder is a Pauli operator on the surface code. The partial correction on a qubit is determined by the parity of all matchings made on space-like edges in the syndrome history corresponding to that qubit. The partial correction can be specified by a bitstring \(x_{(u, v)}\), where non-zero entries of \(x_{(u, v)}\) indicate the presence of a correction. Specifically, we have:

\begin{equation} 
    x_{(u, v)} = \sum_t s_u(t) s_v(t). \label{eq:correction} 
\end{equation} 

One important quantity that characterizes the pre-decoder is its \emph{isolation volume}, which we denote \(V_0\). The pre-decoder will only accurately correct a single error if there are no other errors within a spacetime volume \(V_0\) of the error. By drawing an isolated defect pair and by calculating the effect of other errors on nearby edges, we can compute \(V_0 = 57\), and this is shown in Appendix~\ref{sec:recovering-failure-rates}. If error rates are large such that \(p V_0 \gg 1\), then there are very few isolated defect pairs in the syndrome history and pre-decoding is ineffective. On the other hand, if error rates are sufficiently low that \(p V_0 \ll 1\), then most defect pairs are isolated and pre-decoding can significantly reduce the bandwidth and latency of the scheme. We refer to these as the \emph{high error-rate} and \emph{low error-rate} regimes respectively. Importantly, we remark that the effectiveness of pre-decoding depends only on the error rate, and not on the code size. When pre-decoding does result in savings, these savings do not saturate for large code sizes.


We can now turn our investigation into the costs and savings of using the pre-decoder as the first stage in a two-stage decoding scheme. We will explore the asymptotic QEC performance in Sec.~\ref{sec:failure-rates-and-threshold}, the expected gains in latency and bandwidth in Sec.~\ref{sec:savings}, and finally we will investigate how this pre-decoder is expected to function in practice for fault-tolerant quantum computing in Sec.~\ref{sec:practical-cost}.

\section{Asymptotic performance \label{sec:failure-rates-and-threshold}} 

\begin{figure*} 
    \includegraphics{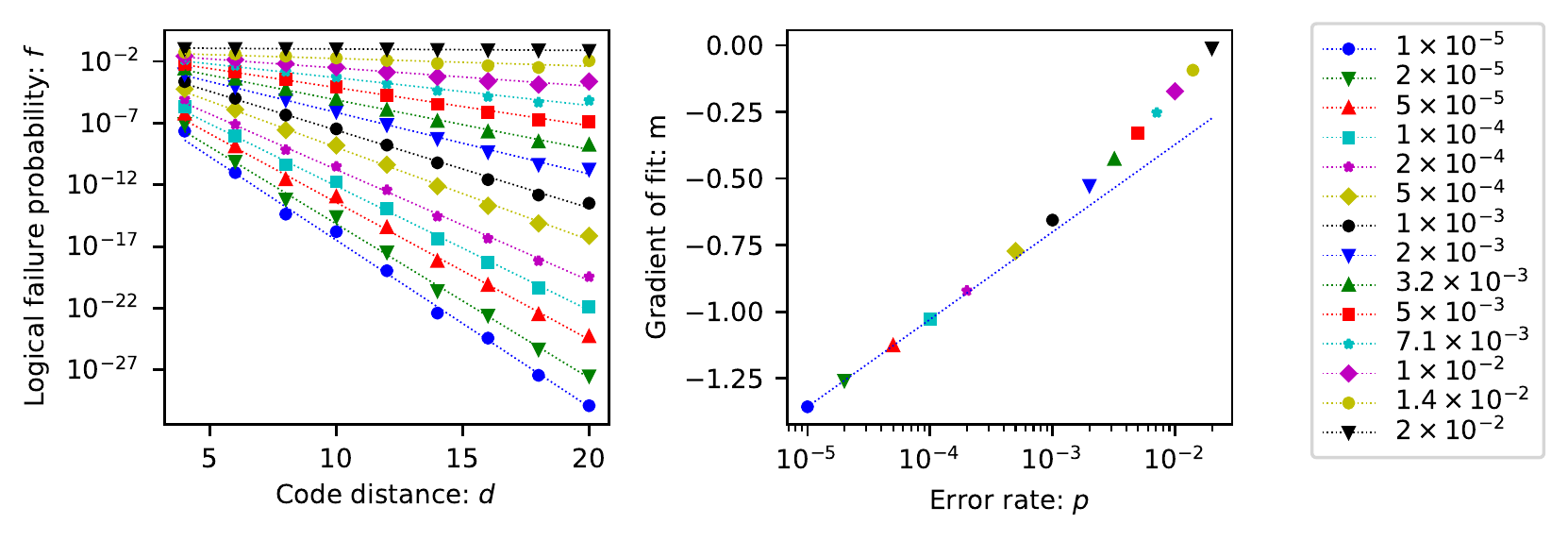}
    \caption{\label{fig:threshold-scaling} (left) Failure probability as a function of code distance when the pre-decoder is used together with MWPM for a range of error rates. We observe that the failure probability decays exponentially in the code distance, indicating the presence of a threshold. (right) Gradients \(m\) extracted from a linear fit \(\log(f) = md\) of the decay of the failure probability, as a function of physical error rate \(p\). For small error rates (\(p \lesssim 10^{-3}\)), we find \(m = k \log(p)\) for \(k = 0.33(1)\).}
\end{figure*} 

In this section, we investigate the expected performance of pre-decoding for large code families.  We first demonstrate that failure probabilities are exponentially suppressed in the code distance for small error rates when pre-decoding is used, indicating the existence of a threshold for the scheme. We also compute the decay constant associated with the sub-threshold scaling of the failure probability that characterizes this suppression in the limit of large codes and small error rates. These should be understood as preliminary results that indicate the viability of the scheme as a FTQEC protocol. A more in-depth analysis of the costs and benefits of the scheme for real-time decoding applications are give in Secs.~\ref{sec:savings} and~\ref{sec:practical-cost}. 

An ansatz for the sub-threshold scaling of the failure probability is given by 
\begin{equation} 
    f = C \qty(\frac{p}{p_\text{th}})^{kd} \label{eq:scaling} 
\end{equation} 
where \(k \) is the decay constant, \(p_\text{th}\) is the threshold and \(C\) is just a fitting constant. This ansatz describes failure probabilities by extrapolating to low error rates from threshold whilst ignoring combinatoric factors.

The exponent \(kd\) described in Eq.~\eqref{eq:scaling} can be understood as the weight of the least-weight error that results in a logical failure under the scheme. We can therefore make predictions for $k$ for the scheme based on the greediness of the matchings made by the pre-decoder. For MWPM without pre-decoding, there exists string-like errors of weight \(d/2\) that can cause the decoder to fail, so we have \(k = 1/2\) in that case. Consider taking such an error and removing a single phase-flip somewhere away from the boundary of the string. This error will now be correctly decoded by the MWPM algorithm. However, if we first run the pre-decoder over such an error, it will see a pair of adjacent defects where the phase-flip was removed and will re-introduce the phase-flip back into the code. If we follow that up with MWPM then the whole scheme will fail. This describes a failing error of weight \(d/2 - 1\).  In fact, we can remove every third phase-flip from the bulk of an error string and the pre-decoder will re-introduce those phase-flips back into the code. This construction is similar to the Cantor set errors observed in hard-decision renormalization group (HDRG)-style decoders~\cite{ImprovedHDRG}, and leads us to predict a decay constant \(k = 1/3\). We also expect that the threshold of the scheme should not suffer drastically from pre-decoding. The threshold error error rate for standard MWPM with phenomenological noise and without pre-decoding is \(p = 2.9\%\)~\cite{MWPMPhenomenologicalThresholdBitFlip}. This error rate, which gives $p V_0 = 1.7$, is in the transition regime between low and high error-rates as defined in Sec.~\ref{sec:predecoding}. We therefore expect that many of the corrections returned by the pre-decoder at this error rate are still accurate. 

We use various numerical methods to extract the fitting parameters \(k, p_\text{th}\) in Eq.~\eqref{eq:scaling}. In this and other simulations in this work, we used the qecsim software package~\cite{qecsim}, along with other scientific~\cite{scipy} and matching software \cite{Kolmogorov2009} to perform numerical computations. For each code distance in the range \(4 \leq d \leq 22\), we performed a direct Monte Carlo simulation to attain a failure probability at the error rate \(p = 0.02\). In each case, the simulation was run until ten logical failures were observed. For smaller error rates, it was not feasible to use direct Monte Carlo simulation because it would take a prohibitively long time to observe any logical failures. Instead, we used the splitting method \cite{SimulationOfRareEvents, RoleOfEntropyInTQEC, EstimationOfFreeEnergyDifferences}. The splitting method allows us to compute the ratio of the failure probabilities at two different error rates by sampling from the set of failing errors at each error rate. We multiply these ratios together to attain logical failure probabilities at very small error rates, and we use the result of the direct Monte Carlo simulation at \(p=0.02\) as a reference. At each error rate, we collected 1000 independent samples via a Metropolis-Hastings algorithm and computed logical failure probabilities at error rates down to \(p=\num{1e-5}\). 

We plot the failure probability against the code distance in Fig.~\ref{fig:threshold-scaling}. The failure probability is exponentially suppressed in the code distance, indicating that the threshold is persistent when the pre-decoder is run in combination with MWPM. We can also determine that the threshold is approximately \(p = 2\%\), since the logical failure probability is independent of code distance at this error rate. We compare this to the known threshold for MWPM under a phenomenological phase flip error model which is 2.9\%~\cite{MWPMPhenomenologicalThresholdBitFlip}. We also extract the decay constant from our simulations in Fig.~\ref{fig:threshold-scaling}, and find \(k = 0.33(1)\), consistent with our analysis above. We contrast this value against the decay constant for MWPM, which is \(k = 0.5\).

\section{Bandwidth and latency \label{sec:savings}} 

In this section we study the effect that pre-decoding has on the defect density in a syndrome history, as well as on the runtime of the main MWPM decoder. As the syndrome is defined in a spacetime volume, the defect density can be thought of as the bandwidth per unit area of the quantum layer required to transmit the syndrome. The defect density also determines the size of the problem that must be solved by the main decoder. We provide the mean savings to the defect density and MWPM runtime that are realized in the low error-rate regime, and additionally provide the full defect count distribution since the worst-case bandwidth usage and runtime are also important quantities of interest for real-time decoding. 

\subsection{Defect count distribution}

We first develop a heuristic model for the distribution of defect counts and treat the defect density as a parameter of this model. Let \(M\) be the random variable corresponding to the number of defects sent to the main MWPM decoder in one syndrome history of spacetime volume \(V\). We do not yet specify whether pre-decoding is being used. We note that neither the noise model nor the pre-decoder introduce any long-range correlations into the syndrome data. If we coarse grain the syndrome history into suitably sized blocks, the total defect count is the sum of many independent binomial random variables corresponding to whether or not there is a defect pair in each block. The total expected number of defects is fixed by the defect density. Since the blocks can be made small, up to a size \(V_0\), the total defect count can be modelled using a Poisson random variable. Technically, since the total defect number is constrained to be even by an emergent symmetry~\cite{KitaevToricCode}, it is really \(M/2\) that is described as a Poisson and we have
\begin{equation} 
    \frac{M}{2} \sim \; \text{Poisson} \qty(\frac{\bar{\rho}{V}}{2}) \label{eq:bw_model} 
\end{equation} 
where the mean defect density \(\bar{\rho}\) depends on the error rate, as well as whether or not pre-decoding is being used. It should be noted that Eq.~\eqref{eq:bw_model} describes the number of defects in a full cycle of fault-tolerant error correction. The distribution that describes the bandwidth usage during each round of measurement is given by replacing \(V\) by \(V/t\), where \(t\) is the number of rounds of stabilizer measurement and we expect to set \(t = d\). 

\subsection{Defect density with and without pre-decoding}

We now study the defect density under MWPM-only and under pre-decoding. For the case where no pre-decoding is performed, or in the high error-rate regime when pre-decoding is not useful, the bandwidth density is proportional to the error rate with a coefficient of two to account for the fact that a single fault creates two defects. We have
\begin{equation} 
    \bar{\rho}_\text{MWPM} = 2 p. \label{eq:density-MWPM}
\end{equation} 
In the low error-rate regime and when pre-decoding is used, the defect density can be significantly reduced. The probability of two errors occuring within a given spacetime volume \(V_0\) is given by \((V_0 p)^2/2\). Most of the error configurations that result in an inaccurate correction by the pre-decoder leave behind two defects in the syndrome history after pre-decoding, and this is shown in Appendix~\ref{sec:recovering-failure-rates}. These considerations lead to a defect density given by
\begin{equation}
    \bar{\rho}_\text{pre} = p^2 V_0. \label{eq:density-pre}
\end{equation}

\begin{figure}
    \includegraphics{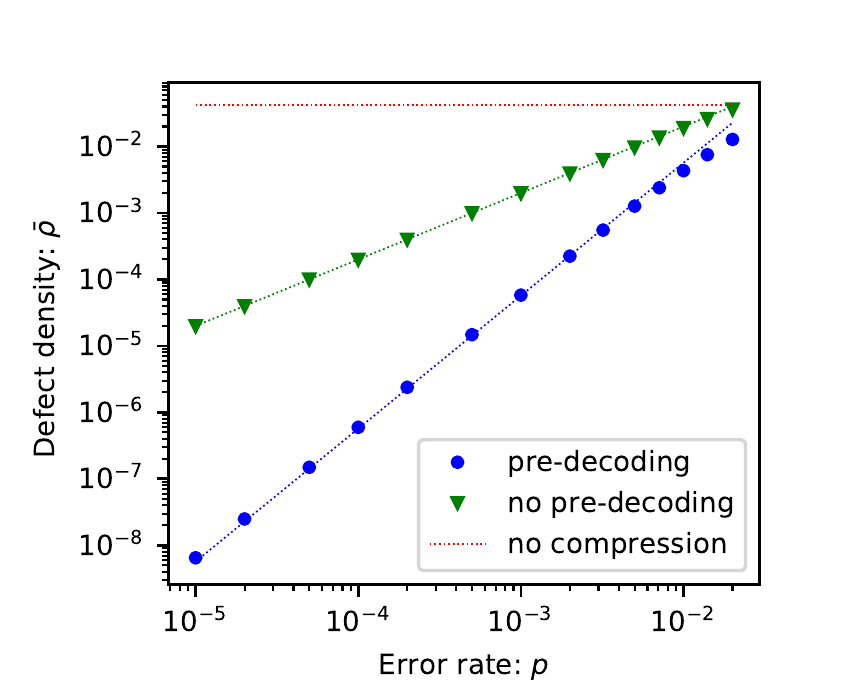}
    \caption{\label{fig:bandwidth_savings}  Defect density as a function of error rate.  The defect density determines the required bandwidth for transmitting syndrome information per unit area.
    When syndrome compression is used (green), the defect density scales linearly in the error rate. When the pre-decoder is also used (blue), the defect density scales quadratically in the error rate. The dotted green line corresponds to Eq.~\eqref{eq:density-MWPM}, and the dotted blue line corresponds to Eq.~\eqref{eq:density-pre} with \(V_0 = 57\). (red) The bandwidth usage associated with an uncompressed syndrome history is independent of the error rate. In order to make a fair comparison, we show density of vertices in the syndrome history divided by 16. The factor 16 assumes that addresses require 16 bits to transmit, whereas measurement data requires only one bit.} 
\end{figure} 

We verified these predictions numerically. We used direct Monte Carlo methods to estimate the defect density over a range of physical error rates when syndrome compression is used, and when syndrome compression and pre-decoding are used together. For each physical error rate, we computed the mean number of defects in the syndrome histories for codes of different sizes, averaged over at least 10000 decoding cycles, with more decoding cycles used for small error rates. We then fit a linear model in order to extract the defect density. In Fig.~\ref{fig:bandwidth_savings}, we observe that the defect density is  reduced by a factor that scales quadratically in the physical error rate, as predicted by Eq.~\eqref{eq:density-pre}. For small error rates \(p = 10^{-4}\), the defect density is reduced by a factor \(10^5\) as compared to when no pre-decoding or syndrome compression is used. At more moderate error rates of \(p = 10^{-3}\), we still see reductions of order \(1000\).

We make two observations about these bandwidth savings. First, the mean bandwidth savings do not depend on the size of the code, because when the pre-decoder encounters a difficult error configuration in the syndrome history, it will only be deterred from performing greedy matchings in a very small neighbourhood of that error. Second, bandwidth savings are made at all error rates below threshold,  because the low error-rate regime is set by \(p = V_0^{-1} = 1.75\%\) and this value happens to be very close to the threshold for this scheme. 

\subsection{Worst-case performance for finite-size quantum computations}

\begin{figure}
    \includegraphics{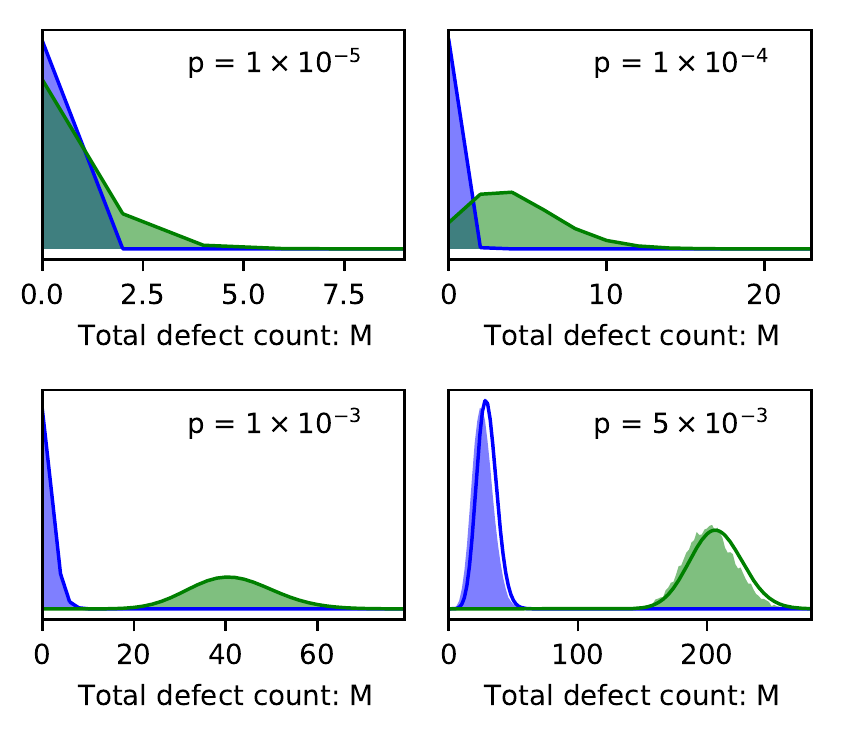}
    \caption{\label{fig:histogram} Histograms of the number of defects contained in a distance-20 syndrome history with pre-decoding (blue) and without pre-decoding (green). The solid blue and green lines correspond to the random variables described by Eq.~\eqref{eq:bw_model} for \(\bar{\rho} = p^2 V_0\) and \(\bar{\rho} = 2p\) respectively.} 
\end{figure} 

In a real device, it may be a requirement that at all times the bandwidth usage does not exceed some maximum value set by device constraints. If the distribution of bandwidth used in a decoding cycle is highly skewed towards large bandwidths, then the mean bandwidth saving is not a good metric for the efficacy of the scheme. Instead, the efficacy depends on the frequency with which large bandwidth-demanding events occur in the system. 

We quantify this by following Ref.~\cite{HierarchicalDecoding} and study the top \((1-f)\)-percentile of the defect count, denoted \(M_\text{max}\) and defined by \(P(M > M_\text{max}) = f\), where we set \(f = 10^{-15}\) as a target logical failure probability. 

We collected data on the full histogram of defect counts generated by the Monte Carlo simulations. In Fig.~\ref{fig:histogram} we compare this data to our heuristic model in Eq.~\eqref{eq:bw_model} for a few select error rates and a spacetime volume corresponding to a distance $d=20$ surface code. 

When the spacetime volume is large, we can appeal to the law of large numbers. In that case, the observed defect count will be close to its mean value with high probability, on a logarithmic scale. For instance, for \(\bar{\rho} V > 1000\), we have that \(M_\text{max} / \bar{\rho} V < 1.4\). In that case, we may freely replace \(\bar{\rho}\) with \(M_\text{max}/V\) in Fig.~\ref{fig:bandwidth_savings}, and the error bars would not be visible.

For small values of \(\bar{\rho}V/2\), the distribution can be seen to be skewed towards high bandwidths, and this will slightly erode the savings to the bandwidth usage that are realized. This effect can be straightforwardly calculated using properties of the Poisson distribution. We emphasize that provided that multiplexed resource-sharing techniques are used \cite{ScalableDecoderMicroarchitecture, HierarchicalDecoding}, the defect count can be averaged over a spacetime volume that corresponds to the syndrome histories for all logical qubits sharing a line of communication with the main decoder. This results in a bandwidth distribution with smaller variation than would be observed for a single logical qubit.  

We remark on the particular applicability of this scheme to performing entangling gates in quantum computations. Certain schemes for performing entangling gates require that  different patches of surface code be fused together to form bigger patches of surface code. It may then be the case that we require fault-tolerant quantum error correction to be carried out over a large surface code patch. This will be difficult if the bandwidth savings saturate at large code sizes due to any global conditionals in the pre-decoding algorithm. The local and greedy nature of the CA pre-decoder is essential to it being able to reduce the density of defects, and it lends itself naturally to this setting.

\subsection{MWPM runtime and latency}

\begin{figure}
    \includegraphics{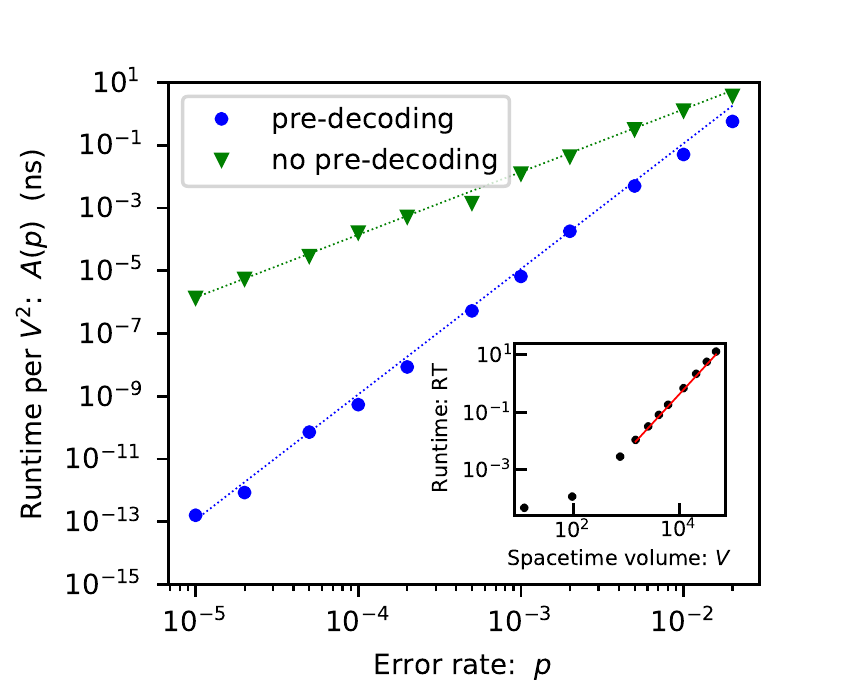}
    \caption{\label{fig:runtime} For each error rate, we plot the constant of proportionality extracted from a fit of runtime \(RT = A(p) V^2\), where \(V\) is the spacetime volume of the syndrome history and the runtime is for the global matching part of the decoding scheme, and is given in nanoseconds. (blue) Pre-decoding is used. Markers are data points, and the dotted line is the fit \(A(p) = A (V_0 p/2)^2 p^2\), for \(A = \num{1.4e-4}\). (green) No pre-decoding is used. Markers are data points, and the dotted line is the fit  \(A(p) = A p^2\) for the same \(A = \num{1.4e-4}\). (inset) we demonstrate the goodness of the fit \(RT = A(p) V^2\) (red) to data (black) for the example case of no pre-decoding at error rate \(p=\num{2e-2}\).}
\end{figure}

The reduction to the defect density translates directly to a speed up in the runtime of the global decoder. In the implementation of MWPM that we use~\cite{Kolmogorov2009}, the worst-case runtime when applied to complete graphs is \(O(|W|^3)\), where \(\abs{W}\) is the number of vertices. However, significant effort has gone into ensuring that this algorithm runs faster in most cases. Matching algorithms based on complete graphs generated by quantum error correction have been reported to have more favourable scaling in practice than their worst-case behaviour would suggest \cite{pymatching}. In the inset to Fig.~\ref{fig:runtime}, we demonstrate that the runtime is well described by \(\text{RT} \sim \abs{W}^2\).

We can determine from Eqs.~(\ref{eq:density-MWPM}) and~(\ref{eq:density-pre}) that the factor speed up in the MWPM decoder is expected to depend on the error rate as
\begin{equation}
    \frac{\text{RT}_\text{pre-decoded}}{\text{RT}_\text{not\ pre-decoded}} = \qty(\frac{p V_0}{2})^2 
\end{equation}
We can understand this speed up by noting that this MWPM implementation requires a complete graph to be constructed for every error configuration (although cf.~\cite{MWPMINOrder1}). This implementation does not exploit the fact that isolated errors are easier to correct than large clusters of errors. The runtime savings from predecoding arise by matching isolated pairs of defects greedily at the start so they do not need to be included in the complete graph. In Fig.~\ref{fig:runtime}, we demonstrate the speed up in the limit where decoding is performed over a large spacetime volume. At error rates \(p = 10^{-3}\), we observe that the MWPM decoder is sped up by a factor of 1900. At error rates \(p = 10^{-4}\), we observe a speed-up by a factor of $3.1 \times 10^5$. 

There are two important caveats to note here. First is that these speed-ups are only fully realised when the spacetime volume of the syndrome history is large. Specifically, the speed-up arises only if there are enough defects in the syndrome history that the runtime of the MWPM decoder is dominated by its asymptotic behaviour. For the implementation of MWPM that we use, this is the case when there are more than 50 defects in the syndrome history. The relevant spacetime volume here is the spacetime volume corresponding to a connected patch of surface code that is being decoded as one logical unit. If there are not more than 50 defects in the syndrome history both with and without pre-decoding, then the runtime savings reported above are not fully realized. At error rate \(p = \num{5e-3}\) and code distance \(d = 30\) (sufficient to reach logical failure probability \(f = 10^{-15}\)) this condition is met. A 64X runtime speedup is realized, and this is the full speed up attainable at that error rate. At smaller error rates, this condition is unlikely to be met when using a reasonably-sized surface code as a memory, but may be met when decoding over a large surface code patch, for example when performing an entangling gate. Nonetheless, at error rates \(p = 10^{-3}\) and code distances \(d = 22\) (sufficient to reach logical failure probability \(f = 10^{-15}\)), a direct comparison of runtimes still revealed a speed-up by a factor of 200.

The second caveat is that the worst-case runtime is going to depend on the tail end of the full defect distribution. This discussion is equally relevant to determining the worst-case runtime for the MWPM under pre-decoding. If the worst-case runtime is significantly larger than the mean runtime, then the mean runtime can be recovered by using multiplexed resource-sharing techniques \cite{HierarchicalDecoding, ScalableDecoderMicroarchitecture}. 

\section{Pre-decoding in practice} \label{sec:practical-cost}

The diminished decay constant \(k = 1/3\) in the logical failure probability scaling of Eq.~\eqref{eq:scaling} associated with the pre-decoder, compared with $k= 1/2$ for MWPM, means that larger codes will be needed to achieve a target logical failure probability.  We note, however, that this scaling coefficient is describing the asymptotic performance and may not be capturing the relative performance of the pre-decoding approach for practical code sizes.  We now investigate the costs associated with using the pre-decoder in regimes that are relevant to fault-tolerant quantum computing.  

\subsection{Practical qubit cost of pre-decoding}

\begin{figure} 
    \includegraphics{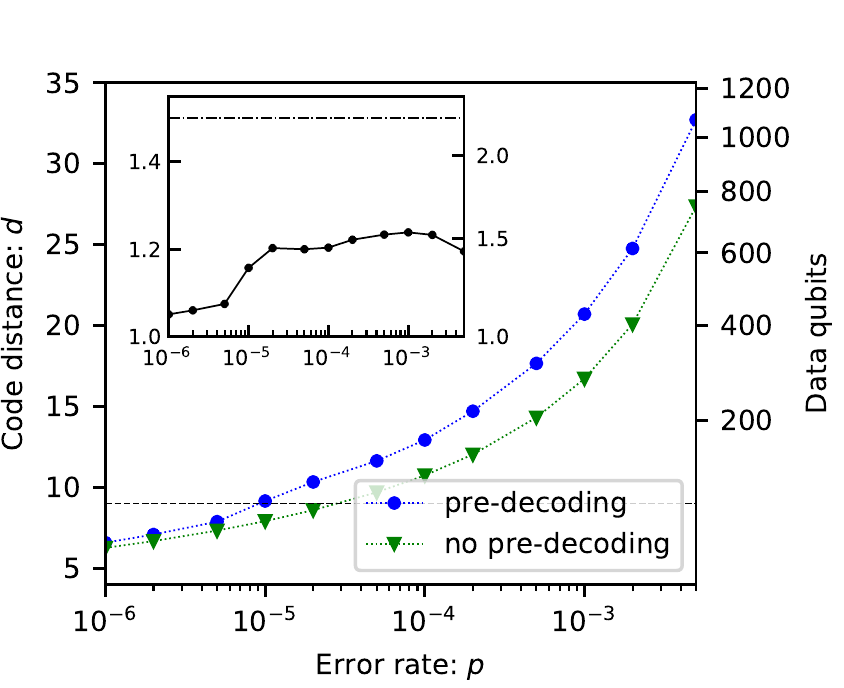}
    \caption{\label{fig:cost_of_predecoding} (main) The code distance (left axis) and number of qubits (right axis) required to reach a target logical failure probability of \(f = 10^{-15}\) when pre-decoding is used (blue) and when pre-decoding is not used (green). Distances are calculating by inverting Eq.~\eqref{eq:sophisticated-scaling}, and using ansatz Eq.~\eqref{eq:weight} and Eq.~\eqref{eq:A_pre},~\eqref{eq:A_no-pre}. The dashed line marks \(d = 9\), below which pre-decoding performs comparably to pure MWPM. (inset) the factor increase in the code distance (left axis) and number of data qubits (right axis) required with pre-decoding compared to without pre-decoding. The dashed-dotted line denotes the expected behaviour using Eq.~\eqref{eq:scaling} with \(k = 1/3\). } 
\end{figure} 

Since Eq.~\eqref{eq:scaling} extrapolates to low error rates from threshold and ignores entropic effects, it may not provide an accurate prediction for logical failure probabilities in all regimes. In Fig.~\ref{fig:cost_of_predecoding}, we compare the code distance required to reach target logical failure probabilities of \(f=10^{-15}\) both with and without pre-decoding. We find that, in order to accommodate the pre-decoding scheme, we must increase the qubit count by no more than 50\%. We also remark that as error rates become very small, there is no qubit cost associated with running the pre-decoder. The decay constant \(k=1/3\) is therefore a pessimistic indicator of performance (it would suggest that the qubit count need be increased by a factor 2.25X). Importantly, we remark that Fig.~\ref{fig:cost_of_predecoding} was generated by a more sophisticated ansatz than that given in Eq.~\eqref{eq:scaling}. We develop our detailed ansatz throughout this section.

\subsection{Structure of the failing error set: weight and multiplicities}

To see why the increase in overhead is modest, we study in detail the structure of the failing error set for small codes, which reveals a more complex resource cost for pre-decoding based on two shortcomings of Eq.~\ref{eq:scaling}.

\begin{figure} 
    \includegraphics{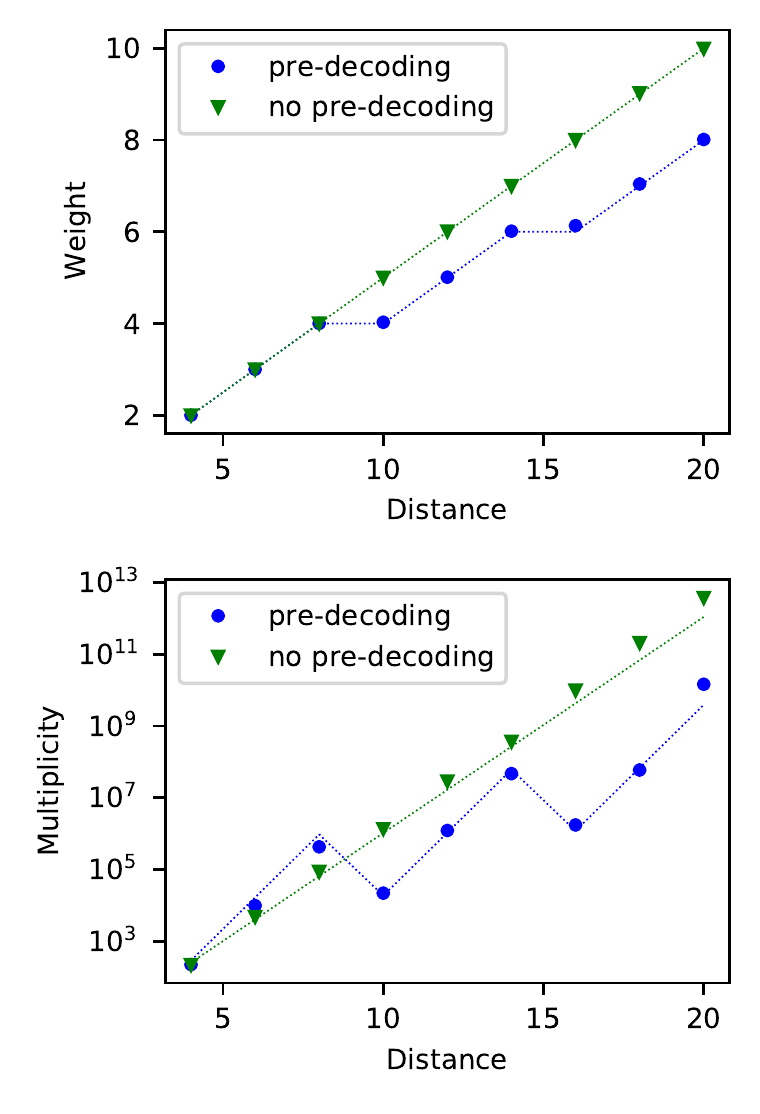} \caption{\label{fig:structure-failing-errors} (top) The weight of the minimum weight failing error, as a function of code distance. The solid lines are analytic and are given by by Eq.~\eqref{eq:weight} for the case of predecoding (blue), or by \(d/2\) for MWPM only (green). Data points are extracted as the gradients of linear fits of the log failure probability to the log error rate (see Eq.~\eqref{eq:sophisticated-scaling}). (bottom) Multiplicity of least-weight failing errors as a function of code distance. Data points are extracted as the intercepts of linear fits of the log failure probability to the log error rate (see Eq.~\eqref{eq:sophisticated-scaling}). The dashed green line is a fit for MWPM of form Eq.~\eqref{eq:A_no-pre}. The dashed blue line is a fit of form \(A(d) \sim 2^{6a - 2 + \alpha b}\), where \(d = 6a + b - 2\) for non-negative integer \(a\) and non-negative integer \(b < 6\) and fitting parameter \(\alpha\). When \(b = 0\), the scaling of \(A(d)\) is described by Eq.~\ref{eq:A_pre}, and this occurs here for distances \(d = 4, 10, 16\). }
\end{figure}

First, we note that Eq.~\eqref{eq:scaling} ignores entropic factors associated with the number of errors that result in a logical failure (see \cite{TopologicalQuantumMemory, RoleOfEntropyInTQEC}). Whilst Eq.~\eqref{eq:scaling} is good for predicting the \emph{scaling} of logical failure probabilities at low error rates, it is not calibrated in all cases to predict logical failure probabilities themselves. In our scheme, at modest system sizes we find that the multiplicity of low-weight errors is relatively small, such that the logical failure probability is not impacted significantly by the leading order scaling due to the contribution from low-weight logical errors. The second shortcoming of Eq.~\eqref{eq:scaling} is that the decay constant to fit Eq.~\eqref{eq:scaling} assumes that the weight of the least-weight failing error is linear in the code distance. In our scheme, this is not true for small codes due to finite size effects. 

To capture these effects, we make use of the fact that the weight and multiplicity of the least-weight failing error often provides much of the crucial information required in QEC for decoding moderately sized codes \cite{RateAgnosticMCMC}. That is, we can construct an ansatz for the logical failure probabilities based on these this data that will be relevant for a practical regime. We denote by \(\lw(d)\) the weight of the least-weight error that results in a logical failure, and we denote by \(A(d)\) the multiplicity of such errors.  We emphasize that both of these quantities depend not only on the code, but also on the decoding scheme that is used. In particular, we have generally that \(\lw(d) \leq d/2\), with the equality satisfied in the case of MWPM without pre-decoding. We will use subscripts where necessary to indicate the decoding scheme. The failure probability of the code is then given:
\begin{equation} 
	f = A(d) p^{\lw(d)}. \label{eq:sophisticated-scaling} 
\end{equation} 
We now investigate the minimum weight of such failing errors, followed by the multiplicity of these errors.

\subsection{Minimum weight of failing error}

If we follow the prescription in Sec.~\ref{sec:failure-rates-and-threshold} for constructing least-weight failing errors, then taking into account finite size effects leads to an ansatz: 
\begin{equation} 
	\lw_\text{pre}(d) = \ceil*{\frac{2}{3}\qty( \frac{d}{2} + 1) } \label{eq:weight} 
\end{equation} 
where \(\lw_\text{pre}(d) = kd\) for \(k = 1/3\) is recovered for large codes. We plot our data with the ansatz in Fig.~\ref{fig:structure-failing-errors} to show the correlation between our model and our results. Since we have generally that \(\lw(d) > d/3\), we remark that \(k = 1/3\) is overly pessimistic for predicting logical failure probabilities. It is particularly significant however that for \(d < 9\) we have \(\lw(d) = d/2\). We therefore expect the predecoder to perform comparably to MWPM for code sizes \(d < 9\), with performance differences arising only from combinatoric factors. In Fig.~\ref{fig:effectiveMWPM}, we plot the code distance required for a MWPM-only decoder to reach the same logical failure probability as when pre-decoding is used. We call this quantity the effective MWPM distance and denote it \(d'\). We can see that the effective MWPM distance approaches the actual code distance for \(d < 9\) and low error rates, indicating that pre-decoding comes at almost no cost in this regime. 

\subsection{Multiplicities of failing errors}

Another important structural feature of the failing error set under pre-decoding is that there are very few least-weight failing errors compared to higher-weight errors. This can be easily visualized by considering some weight \(d\) logical operator on the surface code. A weight \(d/2\) error that will cause a logical failure under MWPM can be constructed by placing a phase-flip on any \(d/2\) data qubits in the support of the logical. There are approximately \({d \choose d/2} \lesssim 2^d\) of making this selection. On the other hand, with pre-decoding in place a weight \(d/3\) error will only result in a logical failure if it is carefully constructed by taking a length \(d/2\) contiguous string of affected qubits, and then removing every third qubit from the support of the error. This means that there are far fewer least-weight failing errors under pre-decoding. 

We can formulate these statements into an ansatz for the multiplicites of failing errors with and without pre-decoding. In both cases, there is a factor \(2^d\) that arises by counting the number of logical operators in the surface code (see \cite{RoleOfEntropyInTQEC} for an explicit calculation of these combinatoric factors). For the case of MWPM, there is an additional factor \(2^d\) due to the arguments above. This gives the rough ansatz for the multiplicites
\begin{gather} 
    A_{\text{pre}}(d) \sim 2^d = 2^{3 \lw_\text{pre}(d) - 2} \label{eq:A_pre}\\
    A_{\text{no pre}}(d) \sim 4^d = 2^{4 \lw_\text{no pre}(d)} \label{eq:A_no-pre} 
\end{gather} 
where we have ignored the ceiling function in the second equality in Eq.~\eqref{eq:A_pre}. We remark that strictly speaking, the above ansatz applies to pre-decoding only on code distances where the ceiling function in Eq.~\ref{eq:weight} can be ignored. In a sense, this captures a situation where we can only just form a failing error of weight \(\lw_\text{pre}(d)\). We give a more sophisticated ansatz in Fig.~\ref{fig:structure-failing-errors} and also compare these ansatz to numerical data.

\begin{figure} 
    \includegraphics{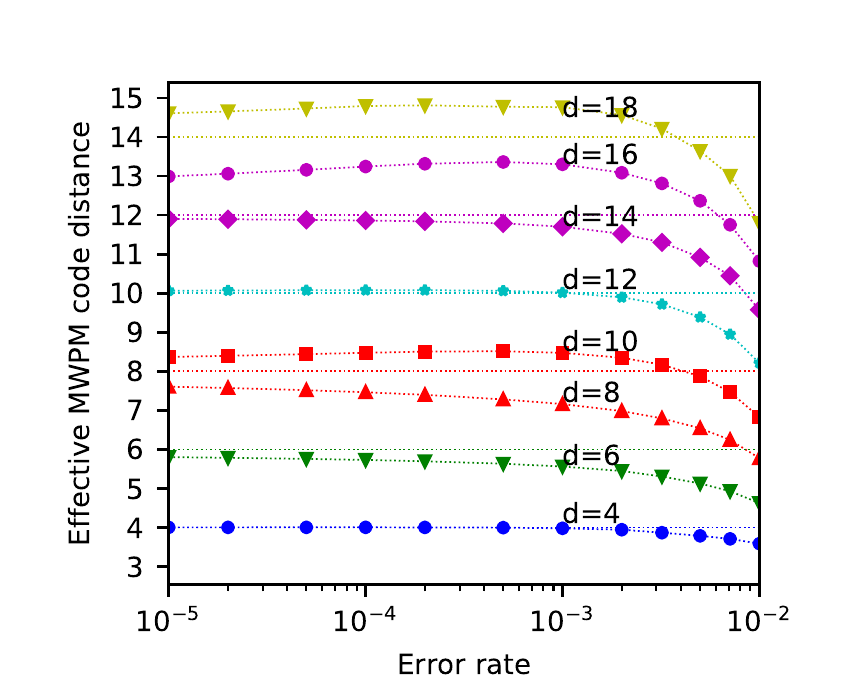}
    \caption{\label{fig:effectiveMWPM} Effective MWPM code distance of the pre-decoding scheme as a function of error rate, plotted for code distances \(d = 4, 6, 8, 10, 12, 14, 16, 18\). Solid lines are color-coded to correspond to dashed lines that provide the low error rate asymptotic behavior, as determined by the weight of the least-weight failing error and given by Eq.~\eqref{eq:weight}.}
\end{figure}

A comparison that is relevant for assessing decoding performance is to fix the weight of the least-weight failing error and then compare the multiplicities under pre-decoding to the multiplicities without pre-decoding. The second equalities above allow us to make that comparison and indicate that the multiplicities are more favourable when pre-decoding is performed for moderately sized codes. The distance-16 surface code under pre-decoding highlights this point. Since we have that \(\lw_\text{pre}(16) = \lw_\text{no pre}(12)\), it would be reasonable to expect that pre-decoding over a distance-16 surface code should perform similarly to pure MWPM over a distance-12 surface code. In fact, from Fig.~\ref{fig:effectiveMWPM}, we see that pre-decoding over a distance-16 surface code outperforms pure MWPM over even a distance-13 surface code due to the favourable multiplicities under pre-decoding. This effect becomes more pronounced for larger code sizes. 

We remark that in a practical regime set by some fixed target logical failure probability, the exact form of \(\lw_\text{pre}(d)\) and the effect of multiplicities are both important to consider. Which effect is more important depends on the underlying error rate. If error rates are low, then large codes are not required and the precise form of \(\lw_\text{pre}(d)\) for small codes must be considered. On the other hand, if error rates are high, then large codes are required. The multiplicities of least-weight failing errors becomes a significant effect that must be accounted for. 

In this and the previous section we have developed numerical methods and ansatz that have allowed us to understand the bandwidth usage and runtime savings resulting frome pre-decoding, as well as the logical failure probabilities. We go on in Appendix~\ref{sec:recovering-failure-rates} to extend this analysis to a parameterized family of pre-decoders that interpolates between the pre-decoder studied here, and an implementation of a standard MWPM decoder, and we detail a trade-off between the resources savings and the accuracy of the pre-decoder.

\section{Discussion \label{sec:discussion}}

We have proposed a two-level decoding architecture for the surface code based on a local pre-decoder that makes greedy matchings designed to clean up sparse errors. Our numerical simulations demonstrate that the pre-decoder reduces the complexity and runtime of the global matching decoder's task, and also delivers substantial savings in the bandwidth density. Remarkably, despite the pre-decoder's intended design for low error rates, we find that these advantages persist up to threshold error rates. These order of magnitude improvements come at a modest increase in the qubit overhead to recover a commensurate logical failure probability compared with global approaches to decoding. Our results therefore show that augmenting our global decoder with a local pre-decoding system can help to address many of the practical challenges involved with producing a large-scale quantum computing architecture. 

We conclude by identifying some future directions for new research in this area. First, it has been observed that quantum error-correcting systems can be markedly improved by specialising a decoder to correct for the errors of a tailored code~\cite{hookErrors, Tuckett2018, XZZX, XZZXPractical}. It will be interesting to find more sophisticated pre-decoders that account for more realistic noise models. Furthermore, we expect that we can design pre-decoders that are sensitive to hook error syndromes~\cite{hookErrors}, or perhaps flag stabilizer readings~\cite{HeavyCodes, ColorCodeTrivalent}, that indicate where a circuit noise model may introduce a correlated error. We also note that a variant of the pre-decoder presented here can be implemented \emph{within} the quantum layer with Toffoli gates and qubit reset, offering the prospect of pre-decoders requiring no mid-circuit measurement or classical processing.

Lastly, it will be interesting to determine the role of pre-decoders as quantum technology develops. In our work, we have proposed a very greedy decoder designed to significantly reduce the bandwidth cost at the expense of logical failure probability. We contrast this with the approach of Ref.~\cite{HierarchicalDecoding}, wherein a fast all-or-nothing decoder was studied that looks for a globally optimal correction for certain errors, and calls for support from the global decoder where this is not possible. Where the latter approach maintains a better logical failure probability, the former demonstrates good performance for minimising the bandwidth cost, and particularly for noisier near-term devices. Ideally, we may want to balance these considerations, and in Appendix~\ref{sec:recovering-failure-rates} we study a method to interpolate between these two types of pre-decoders. Given that recent experiments have demonstrated classical feedforward together with mid-circuit measurements~\cite{HoneywellRealTime, QECGoogleAI, IBMRealTime, ETHZurich}, we are already in a position to begin to design pre-decoder experiments with real experimental hardware, to begin addressing these questions.

\begin{acknowledgements}
This work is supported by the Australian Research Council via the Centre of Excellence in Engineered Quantum Systems (EQUS) project number CE170100009, and by the ARO under Grant Number: W911NF-21-1-0007. Access to high-performance computing resources was provided by the National Computational Infrastructure (NCI Australia), an NCRIS enabled capability supported by the Australian Government, and the Sydney Informatics Hub, a Core Research Facility of the University of Sydney. BJB is grateful for the hospitality of the Center for Quantum Devices, at the University of Copenhagen where parts of this work were completed. BJB is now affiliated to IBM Quantum.
\end{acknowledgements}

\appendix 

\section{Recovering the subthreshold failure rate \label{sec:recovering-failure-rates}} 

If the performance reduction associated with using the pre-decoder is deemed too costly, then it is possible to trade off decoding accuracy against bandwidth and runtime savings by reducing the greediness of the matchings made by the pre-decoder.  We do this by requiring that the pre-decoder check that a particular defect pair is sufficiently well-isolated from any other defects in the code before performing a correction. 

\begin{figure*}

\includegraphics[width=0.9\textwidth]{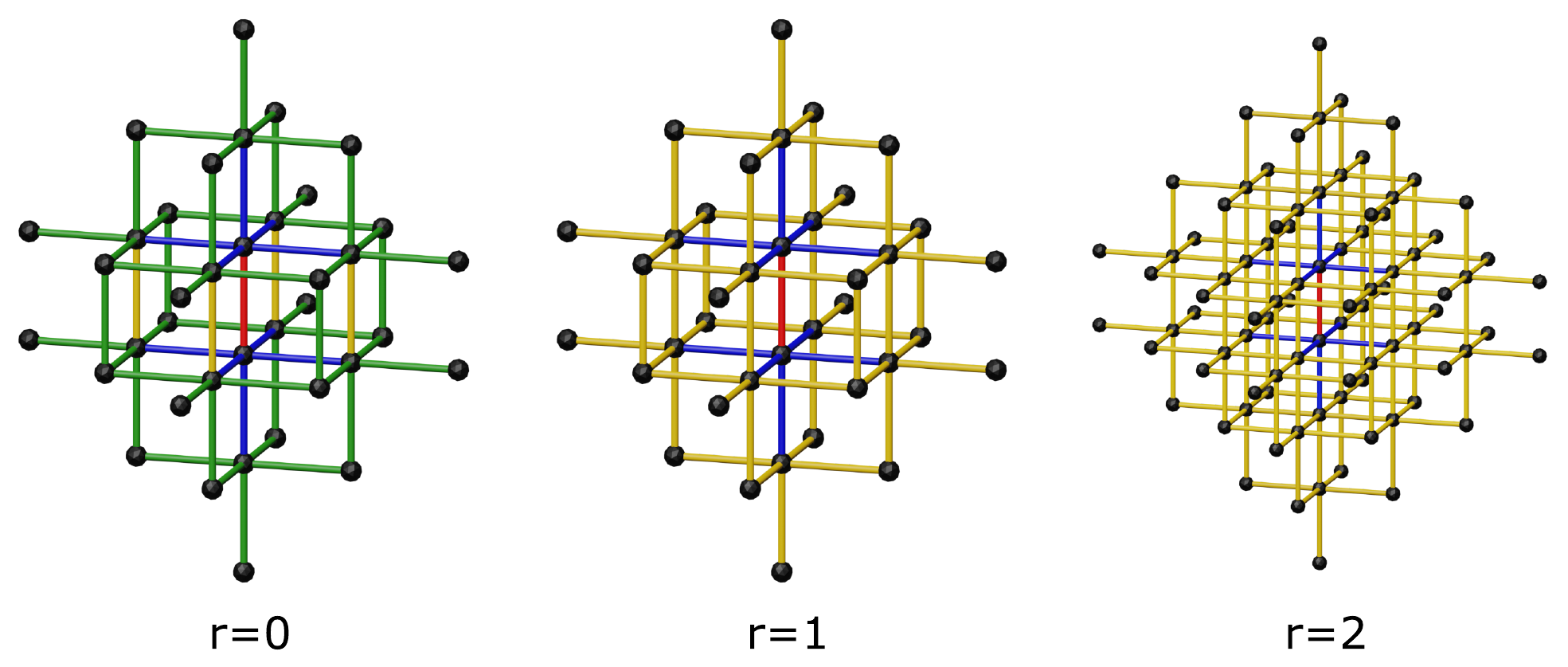}

\caption{ \label{fig:isolation_volumes} Isolation volume of the pre-decoder for \(r = 0, 1, 2\), in order from left to right. Black balls correspond to vertices of the syndrome history, and sticks correspond to edges. In each case, the red stick is the location of a measurement error that has occurred, and other sticks are locations of possible errors that may prevent the pre-decoder from performing an accurate correction. For the \(r=1\) and \(r=2\) pre-decoder (middle, right), an error on any yellow stick will prevent the pre-decoder from performing a correction, and four defects will be left behind in the syndrome history. An error on a blue stick will result in an extended string-like error that is not corrected, and two defects are left behind in the syndrome history. Counting the sticks in the figures gives \(V_1 = 57\) and \(V_2 = 163\). The \(r=0\) case (left) must be treated slightly differently. In this case, the pre-decoder always matches neighbouring defects. If an error has occurred at a blue or yellow stick, the pre-decoder performs either no correction or a correction along a loop which has no effect, leaving behind two or four defects respectively. If an error has occurred on a green stick, then the pre-decoder will return a correction on the green stick, the red stick, and a blue stick that connects the two. This leaves behind two defects in the lattice. In this case, we have \(V_0 = V_1 = 57\), but since here most weight-2 errors result in only two defects left in the lattice, we have that Eq.~\eqref{eq:bw_density_r} contains an additional factor 2 as compared to Eq.~\eqref{eq:density-pre}.} 

\end{figure*} 

More formally, we introduce a family of pre-decoders parameterized by a radius \(r\), representing this isolation distance. This family interpolates between the case \(r = 0\), which was studied in the main text, and the case \(r=\infty\), which is an implementation of the standard global MWPM decoding algorithm. To define this family, we introduce a projector that is equal to one whenever a particular defect in a defect pair is well-isolated:
\begin{equation} 
    \Pi_r(v, t) =
    \begin{cases} 
        1& \text{if } \sum_{(u, t') \in B_r(v, t)} s_u(t') \leq 2 \\
        0& \text{otherwise} 
    \end{cases} 
\end{equation} 
where \(B_r(v, t)\) is a ball with centre \((v, t)\) and radius \(r\) on the decoding lattice, defined with respect to the taxicab metric. Then, writing \(\tilde{s}_v(t) = \Pi_r(v, t) s_v(t)\), we have that the action of the radius-\(r\) pre-decoder results in a modified syndrome: 
\begin{equation} 
    s_v'(t) = s_v(t) + \tilde{s}_v(t) \qty( \sum_{u \sim v} \tilde{s}_u(t) + \tilde{s}_v(t + 1) + \tilde{s}_v(t - 1) ). 
\end{equation} 
Further, the total correction is specified by a bitstring as in Eq.~\eqref{eq:correction}, which is now given as 
\begin{equation} 
    x_{(u, v)} = \sum_t \tilde{s}_u(t) \tilde{s}_v(t) .
\end{equation} 

We roughly sketch how these rules can be implemented using CA with only local communications. The CA will attempt to correct errors according to the rules developed in Sec.~\ref{sec:predecoding}, with a modification. When a CA sees a defect at a particular time, it sends out its address to nearby CA up to a distance \(r\) from itself. With this message passing in place, a CA will not immediately make a correction when it sees a pair of neighbouring defects. Instead, it will wait \(r\) syndrome extraction cycles in order to gather more information. If in that time it receives a signal indicating some nearby defect other than the defect pair being corrected, it will not make the correction. These CAs require some additional memory and processing overheads in order to store and propagate addresses. 

The key parameter that characterizes a pre-decoder in this family is its isolation volume, denoted by \(V_r\). This determines both the accuracy of a given pre-decoder, as well as its bandwidth and runtime savings in terms of a reduction to the defect density. The isolation volume can be computed as in the main text for any \(r\), and we show this in Fig.~\ref{fig:isolation_volumes}. We fit a cubic polynomial to this data for \(r = 1, 2, 3\) and derive
\begin{equation} 
    V_r = 4(\tilde{r} + 1)^3 + 6(\tilde{r} + 1)^2 + 1 \label{eq:isolation_volume}
\end{equation} 
where \(\tilde{r} = \max(r, 1)\). We remark that \(\tilde{r}\) appear instead of \(r\) because there is a sense in which the \(r=0\) pre-decoder is singular in this family of predecoders, as discussed in Fig.~\ref{fig:isolation_volumes}.

\begin{figure} 
    \includegraphics{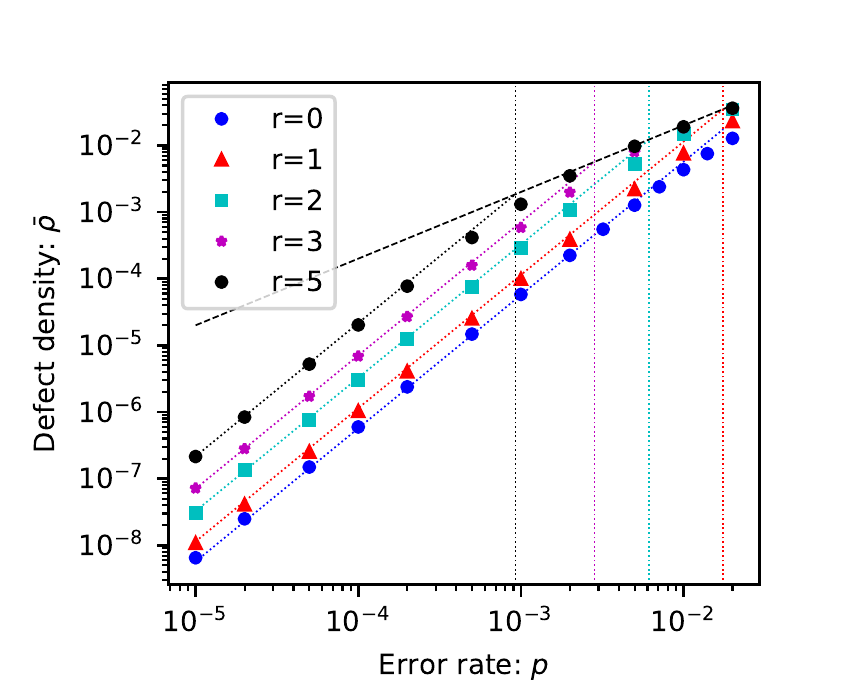}
    \caption{\label{fig:bandwidth_r_nonzero}The defect density associated with pre-decoders for \(r = 0, 1, 2, 3, 5\). The black dashed line corresponds to MWPM-only with syndrome compression. The markers are data points, and the dotted lines correspond to Eq.~\eqref{eq:density-pre} for \(r=0\), and to Eq.~\eqref{eq:bw_density_r} for \(r > 0\), where \(V_r\) is determined by Eq.~\eqref{eq:isolation_volume}. All predecoders exhibit bandwidth scaling with \(p^2\) for sufficiently low-error rates. Vertical lines can be traced to the bottom axis to determine the error rate at which bandwidth savings take effect for a given \(r\).} 
\end{figure}

We follow similar methods as in the main text to study the distribution of defect count under the scheme for different values of the isolation radius \(r\). The distribution is Poisson distributed, as in Eq.~\ref{eq:bw_model}, however the defect density that appears now depends on the isolation radius. In particular, if \(r > 0\), we replace Eq.~\ref{eq:density-pre} with
\begin{equation} 
    \bar{\rho}_\text{pre}(r) = 2 \rho^2 V_r \label{eq:bw_density_r}
\end{equation} 
We remark that the defect density scales quadratically in the error rate for all values of \(r\), provided that the error rate is low enough. However, since Eq.~(\ref{eq:bw_density_r}) is only valid for \(p < 1/V_r\), the error rate at which these bandwidth savings take effect is dependent on \(r\). As we increase \(r\), the pre-decoder is more cautious and requires defect pairs to be more isolated before it will make a greedy matching. This means that error rates must be lower before the defect density is reduced. We compare this model to our numerical data in Fig.~\ref{fig:bandwidth_r_nonzero}. 

\begin{figure} 
    \includegraphics{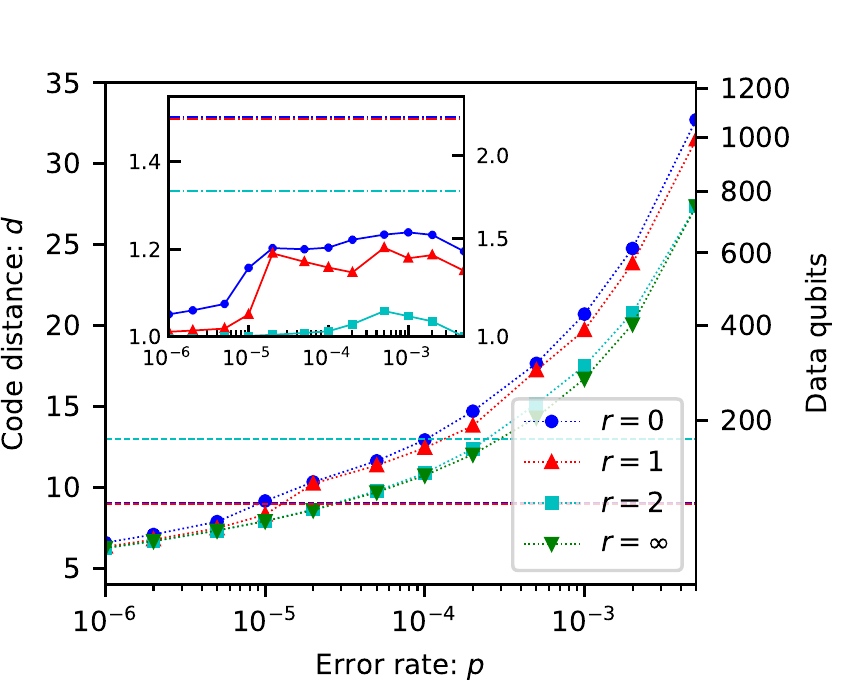}
\caption{\label{fig:cost_of_predecoding_appendix} (main) The code distance (left axis) and number of qubits (right axis) required to reach a target logical failure probability of \(f = 10^{-15}\) for pre-decoders of radius \(r = 0, 1, 2, \infty\), where \(r=\infty\) indicates that no pre-decoding was performed. Distances are calculating by inverting Eq.~\eqref{eq:sophisticated-scaling}, and using ansatz Eq.~\eqref{eq:weight-appendix} along with a similar ansatz to that described in Fig.~\ref{fig:structure-failing-errors} for the multiplicities. These ansatz are linear fits to the logical failure probability at low error rates. The dashed lines are color-coded and mark the maximum distance such that \(\lw_\text{pre, r}(d) = d/2\), below which the pre-decoder of radius \(r\) are expected to perform comparably to pure MWPM. (inset) the factor increase in the code distance (left axis) and number of data qubits (right axis) required with pre-decoding compared to without pre-decoding. The dashed-dotted lines are color-coded and mark the expected behaviour using Eq.~\eqref{eq:scaling} with \(k = (r+1)/(r+2)\).}
\end{figure}

The increased caution taken by the pre-decoder for \(r>0\) also means that it is less likely to introduce erroneous corrections that cause the scheme to fail. For instance, a least-weight error that will cause the pre-decoder of radius \(r\) to fail can be constructed by taking a weight \(d/2\) contiguous error that fails under pure minimum-weight perfect matching and removing every \((r+1)/(r+2)\) phase flip. We generalize Eq.~\eqref{eq:weight} and write
\begin{equation}
    \lw_\text{pre, r}(d) = \ceil*{ \frac{\tilde{r} + 1}{\tilde{r} + 2} \left( \frac{d}{2} + 1 \right) } \label{eq:weight-appendix}
\end{equation}
where \(\tilde{r} = \max(r, 1)\). The qubit cost associated with running the pre-decoding scheme will therefore decrease with \(r\). In the limit of large code sizes and ignoring finite size effects, we have that the weight of the least-weight failing error is linear in the code distance, with decay constant \(k = (r+1)/2(r+2)\). For large \(r\), the pre-decoder makes very few greedy matchings, and the decay constant approaches the \(k=1/2\) that characterizes failure probabilites for MWPM. Of course, for pre-decoding in practice, finite size and combinatoric effects are important, and this can be seen in Fig.~\ref{fig:cost_of_predecoding_appendix}. In particular, we see that for the \(r=2\) pre-decoder and for code distances \(d < 13\), there is almost no qubit cost associated with running the pre-decoder at low error rates. 

Real devices are subject to multiple constraints simultaneously, and we may want to reduce the defect density to some maximum allowed value whilst keeping the logical failure probability as low as possible. This family of pre-decoders allows this trade-off by interpolating between pure MWPM, which provides a high decoding accuracy, and the maximally greedy \(r=0\) pre-decoder, which provides large savings to the bandwidth usage and runtime. 

\end{document}